\documentstyle[preprint,tighten,aps]{revtex}
\begin{document}
\preprint{CERN-TH/97-47~~ROM2F/97/10}
\draft
\title{CP Violating $B$ Decays in the Standard Model and
Supersymmetry} 
\author{M. Ciuchini\thanks{On leave of absence from INFN, Sezione
Sanit\`a, V.le Regina Elena 299, Rome, Italy.}}
\address{Theory Division, CERN, 1211 Geneva 23, Switzerland.}
\author{E. Franco and G. Martinelli}
\address{Dip. di Fisica, Univ. ``La Sapienza"  and INFN,
Sezione di Roma, P.le A. Moro, I-00185 Rome, Italy.}
\author{A. Masiero}
\address{SISSA -- ISAS, Trieste and Dip. di Fisica, Universit\`{a} di Perugia
and INFN, Sezione di Perugia, Via Pascoli, I-06100 Perugia, Italy.}
\author{L. Silvestrini}
\address{Dipartimento di Fisica, Universit\`a di Roma ``Tor Vergata''
and INFN, Sezione di Roma II, 
Via della Ricerca Scientifica 1, I-00133 Roma, Italy.}
\maketitle

\begin{abstract}
We study the uncertainties of the Standard Model (SM) predictions for CP
violating $B$ decays and investigate
where and how supersymmetric (SUSY) contributions may be
disentangled. The first task is accomplished by letting the relevant matrix
elements of the effective Hamiltonian vary within  certain ranges. The SUSY
analysis makes use of a formalism which allows to obtain
model-independent  results. We show that in some cases it  is possible a)
 to measure the CP $B$--$\bar{B}$
mixing phase and b) to discriminate the SM and SUSY contributions to the CP
decay phases.  The golden-plated decays to this purpose are the $B \to \phi K_S$
and $B \to K_S \pi^0$ channels.
\end{abstract}

Searches for CP violation in $B$ decays represent the new frontier in the
realm of Flavour Changing Neutral Current (FCNC) in the Standard
Model (SM) and beyond. SM predictions, however, are plagued
by large uncertainties which have to be taken into account in order to 
probe the SM itself and to disentangle SM effects from
new physics.  A critical assessment of
these uncertainties constitutes a major goal of this work in which we discuss
several possibilities of  looking for signals of low-energy
 supersymmetry (SUSY) in CP violating   $B$ decays.

Early works on FCNC and CP violation were  focused on  a
particular realization of SUSY denoted as the Minimal Supersymmetric 
Standard Model
(MSSM). Remarkably enough, it was realized that the MSSM succeeds to pass all
the challenging FCNC- and CP-tests unscathed. 
This statement holds 
true particularly for CP violation, as long as one puts to zero (or takes very 
small) the two CP violating phases that the MSSM exhibits in addition to the 
usual CKM one. For CP  violation in $B$ physics, under these conditions, the 
MSSM does not yield   major deviations from what we expect in the SM . 

Our view on low-energy SUSY has considerably changed in these last years in 
relation with new insights on the ``parent" N=1 supergravity theories. 
As soon as we move from the MSSM to SUSY-GUT's, or to 
models without universality in the SUSY soft-breaking sector,
we encounter  major differences in    FCNC and CP violating processes. 
In view of  the large variety of low-energy SUSY models that  can be obtained 
by  varying the  ``initial conditions" at some superlarge scale,
it is appropriate to study SUSY predictions as model-independently 
as possible. To this aim, following the early 
work of ref.~\cite{mins}, 
systematic analyses of FCNC phenomena in general SUSY models have been
performed~\cite{fcnc,ggms}. For  CP violation
these studies have  focused so far on $\varepsilon$, 
$\varepsilon^\prime/\varepsilon$ and the electron and
neutron electric-dipole moments.

The impact of new physics on $B$--$\bar{B}$ mixing has been widely explored
(see ref.~\cite{rattazzi} for a review) and SUSY
contributions to  CP phases in $B$-decay amplitudes have recently been
analyzed~\cite{worah}.	When  considering SUSY as an 
example for new physics, these authors, however,
rely  on   some specific SUSY realization. 
Our approach, instead,  allows to draw conclusions which apply to any 
low-energy SUSY extension of the SM. 

This letter addresses two basic and related questions: i) how
large the uncertainties of the SM predictions for CP asymmetries in $B$ decays
 are and ii) in which processes and how one can possibly distinguish SUSY
from SM contributions without making any commitment to a particular model.

Concerning point i), we will work in the theoretical framework of
ref.~\cite{cfms}. We use the effective Hamiltonian
(${\cal H}_{eff}$) formalism, including
LO QCD corrections;  in the numerical analysis, we use the LO SM Wilson
coefficients evaluated at $\mu=5$ GeV, as given in ref.~\cite{zeit}. 
In most  of the cases, by choosing different scales (within a resonable range) 
or by using NLO Wilson coefficients,  the
results vary by about $20-30 \%$ .  This is true
with the  exception of some particular channels where uncertainties  are 
larger. The matrix elements of the operators of ${\cal H}_{eff}$
are given in terms of the following Wick contractions 
 between hadronic states: Disconnected
Emission ($DE$), Connected Emission ($CE$), Disconnected Annihilation ($DA$),
Connected Annihilation ($CA$), Disconnected Penguin ($DP$) 
and Connected Penguin
($CP$) (either for left-left ($LL$) or for left-right ($LR$)
current-current operators). Following ref.~\cite{cfms}, where
a detailed discussion can be found, 
instead of  adopting a specific model for  estimating  the different
diagrams,  we let them vary within reasonable 
ranges.  In order to illustrate the relative 
strength and variation of the different contributions, in table~\ref{tab:ampli} 
we only show, for six different cases,
 results obtained by taking the extreme values of these ranges.
 In the first column 
only $DE=DE_{LL}=DE_{LR}$ are assumed to be different from zero. 
For simplicity, unless stated otherwise,  the same numerical 
values are used for  diagrams corresponding to the insertion 
of  $LL$ or  $LR$ operators, i.e. $DE=DE_{LL}=DE_{LR}$, 
$CE=CE_{LL}=CE_{LR}$, etc.  We then consider, 
in addition to $DE$,  the $CE$ contribution  by taking
 $CE=DE/3$. 
Annihilation diagrams  are included in the third
column, where we use  $DA=0$ and $CA=1/2 DE$~\cite{cfms}. 
Inspired by kaon decays, we allow for some enhancement of 
the matrix elements  of left-right (LR) operators and choose
$DE_{LR}=2 DE_{LL}$ and $CE_{LR}=2 CE_{LL}$  (fourth column). 
Penguin contractions, $CP$ and $DP$, 
can be interpreted  as  long-distance penguin contributions to the matrix 
elements and play an important role:  if   we  
take $CP_{LL}=CE$ and $DP_{LL}=DE$ (fifth column), in some decays 
these terms  dominate the amplitude. 
Finally, in the sixth column,  we allow for long distance
effects which might differentiate penguin contractions with up and charm quarks
in the loop, giving rise to incomplete GIM cancellations (we assume 
$\overline{DP}= DP(c) - DP(u) =  DE/3$ and 
$\overline{CP}= CP(c) - CP(u) =CE/3$). 
For any given decay channel, whenever two terms 
with  different CP phases contribute in the SM, we  give in 
the first row of table~\ref{tab:ampli}, the ratio $r$ of
the two amplitudes.
 
As for the SUSY contribution (point ii)), we make use of the 
parameterization of the SUSY FCNC and CP 
quantities  in the framework of the so-called mass insertion 
approximation~\cite{mins}. For the fermion and sfermion 
states, we choose a basis  where all the couplings of these particles to neutral gauginos are 
flavour diagonal, while the FC arises from the non-diagonality of the sfermion 
propagators. These propagators can be expanded as a series in terms of the 
quantities $\delta \equiv \Delta/m^2_{\tilde q}$, where $m_{\tilde q}$ is an 
average sfermion mass and $\Delta $ denote off-diagonal terms in the sfermion 
mass matrices (i.e., the mass terms relating sfermions with the same electric 
charge, but different flavour). As long as $\delta \lesssim 1$, by taking the 
first term in this expansion   the experimental information on the 
FCNC and CP violating phenomena translates into upper bounds on these 
$\delta$s. Even when the expansion fails, for  $\delta\sim 1$, 
the mass-insertion approximation can still be used as an estimate of the
SUSY effects. 
\par
Using this basis it is possible to account for both gluino- (or neutralino-) 
and chargino-mediated FCNC and CP violation. In view of the 
complexity of the  analysis which includes  chargino contributions, and given 
that the main features are already present with gluinos only \cite{parti},
in this letter we limit 
ourselves  to the SUSY source of CP violation    arising from gluino exchanges.

Four different $\Delta$ mass-insertions in 
the down-squark propagators give rise to  $b \to s $ or $b \to d$ 
transitions: $\left( \Delta_{i3} \right)_{LL}$, 
$\left( \Delta_{i3} \right)_{RR}$, $\left( \Delta_{i3} \right)_{LR}$ and
 $\left( \Delta_{i3} \right)_{RL}$. The indices $L$ and $R$ refer to the 
helicity of the fermion partners. The index $i$ takes the value 1 or 2 
for $b \to d$ or $b \to s$ transitions, respectively. In the
present analysis, we make explicit use  of $\Delta_{LL}$ insertions
only.
While $\left\vert \left( \delta_{23}\right)_{LL}\right\vert$ is left
essentially unconstrained by $b \to s \gamma $, $\left(
\delta_{13}\right)_{LL}$ has to satisfy the bound 
$\left\vert {\rm Re}
 \left( \delta_{13}\right)_{LL}^2\right\vert^{1/2}<0.1 \,  m_{\tilde
q} \, ({\rm GeV})/500$ for degenerate squarks and gluino~\cite{fcnc}. 
 In the following, we
will take $\left\vert\left( \delta_{23}\right)_{LL}\right\vert=1$
 (corresponding to  $x_s = (\Delta M /\Gamma)_{B_s} > 70$   for
the same values of SUSY masses), with amplitudes scaling linearly with 
$\left\vert\left( \delta_{23}\right)_{LL}\right\vert$. 

New physics changes SM predictions on  CP asymmetries in $B$ decays in
two ways: 
by shifting the phase of the $B_{d}$--$\bar{B}_{d}$ mixing  amplitude
and by modifying both  phases and absolute values of  the decay ones. 
The generic SUSY extension of the SM considered here affects all these 
quantities. 

In the SUSY case, by using for the Wilson coefficients in eq.~(12) 
the results of ref.~\cite{ggms} 
and by parameterizing the matrix elements  as we did for the SM case 
discussed above,  we obtain the ratios of  SUSY to SM
 amplitudes given in table~\ref{tab:ampli}. For each decay channel we
give results  for squark and gluino masses of 250 and 500 GeV (second and
third  row, respectively). From the table, 
one concludes that the inclusion of the various terms
in the amplitudes, $DE$, $DA$, etc., 
can  modify the ratio $r$ of  SUSY to SM contributions up to one
order of magnitude.

In terms of the decay amplitude $A$, the CP asymmetry reads 
\begin{equation}
{\cal A}(t) = \frac{(1-\vert \lambda\vert^2) \cos (\Delta M_d t )
-2 {\rm Im} \lambda \sin (\Delta M_d t )}{1+\vert \lambda\vert^2} 
\label{eq:asy}
\end{equation}
with $\lambda=e^{-2i\phi^M}\bar{A}/A$. 
In order to be able to discuss the results  model-independently,
we have labelled as $\phi^M$ the generic  mixing phase.
The ideal case occurs when  one  decay 
amplitude only appears in (or dominates)
a decay process: the CP violating asymmetry is  then determined by the
 total phase   $\phi^T=\phi^M+\phi^D$, where $\phi^D$  is the weak phase
  of the decay.
This ideal situation is spoiled by the presence of
several interfering amplitudes.
If the ratios $r$ in table~\ref{tab:ampli} are small, then the uncertainty on 
the sine of the CP phase is $\lesssim  r $, while if $r$ is O(1)  $\phi^T$ receives,
in general,  large corrections.
\par 
The results of our analysis are summarized in table~\ref{tab:results}.
In the third column, for each channel, we give  the possible SM decay phases  
when one or two decay amplitudes contribute, and the range of variation of their
ratio,  $r_{SM}$,   as deduced from table~\ref{tab:ampli}.
A few comments are necessary at this point:
a) for  $B \to K_S \pi^{0}$ the
 penguin contributions (with a vanishing phase) dominate over the
tree-level amplitude   because the latter is Cabibbo suppressed; 
b) for the channel $b
\to s \bar s d$  only penguin operators or penguin contractions of
current-current operators  contribute; c) the phase $\gamma$ is present in the
penguin contractions of the $(\bar b u)(\bar u d)$ operator, 
denoted as $u$-penguin $\gamma$
in table~\ref{tab:results}~\cite{Fleischer}; d)  
 $\bar b d \to \bar q q $ indicates processes occurring via annihilation 
 diagrams which can be measured
 from the last two channels of table~\ref{tab:results};
e) in the case $B \to K^{+} K^{-}$ both
current-current and penguin operators contribute; f) in $B \to D^{0} \bar
D^{0}$ the contributions
from  the $(\bar b u) (\bar u d)$ and the   $(\bar b c) (\bar c d)$
current-current operators   (proportional to the phase $\gamma$) 
tend to cancel out. 

SUSY contributes to the decay amplitudes with  phases 
induced by  $\delta_{13}$ and
$\delta_{23}$ which we denote as $\phi_{13}$ and $\phi_{23}$. The ratios of
$A_{SUSY}/A_{SM}$ for SUSY masses of 250 and 500 GeV as obtained from 
table~\ref{tab:ampli} are reported in the $r_{250}$ and $r_{500}$ columns 
of table~\ref{tab:results}. 

We now draw some conclusions from the results of table~\ref{tab:results}. 
In the SM, the first
six  decays  measure directly the mixing phase $\beta$, up to
corrections which, in most of the cases, are expected to be small. 
These corrections, due to the presence of  two 
amplitudes contributing with different phases,  produce 
 uncertainties of $\sim 10$\% in   $B \to K_S \pi^{0}$,
 and  of $\sim 30$\%  in $B \to D^{+} D^{-}$ and $B \to
J/\psi \pi^{0}$.   In spite
of the uncertainties,  however, there are cases where
 the SUSY contribution gives rise to significant changes. 
 For example, for SUSY masses of O(250) GeV, SUSY corrections  can  shift the
measured value of the sine of the phase in
 $B \to \phi K_S$ and in $B \to K_S \pi^{0}$ decays by an amount of 
 about 70\%.  For these decays  SUSY effects are sizeable even for 
masses of 500 GeV.  In $B \to
J/\psi K_S$  and $B \to \phi \pi^0$ decays, SUSY effects are only  about $10$\%
but SM uncertainties are negligible.  In $B \to K^0 \bar{K}^0$ 
the larger  effect, $\sim 20$\%,   is partially covered by the 
indetermination of 
about $10$\%  already existing in the SM. 
Moreover the rate for this channel is expected to be rather small.
In $B \to D^{+} D^{-}$  and $B \to K^{+} K^{-}$, SUSY effects are 
completely obscured  by the errors in the estimates of the SM amplitudes.
In $B^0\to D^0_{CP}\pi^0$ the asymmetry  is sensitive to the mixing angle 
$\phi_M$ only because the decay amplitude is unaffected by SUSY. 
This result can be used in connection with $B^0 \to K_s \pi^0$, since
a difference in the measure of the phase  is  a manifestation
of SUSY effects.
\par 
Turning to $B \to \pi \pi$ decays, both the uncertainties
in the SM  and  the SUSY contributions are very large. Here we
witness the presence of three independent amplitudes with different phases 
and of comparable size.  The observation of SUSY effects in
the $\pi^{0} \pi^{0}$ case is hopeless. The possibility of 
separating SM and SUSY contributions  by using the isospin
analysis remains an open possibility which deserves further investigation.
For a thorough discussion of the SM uncertainties in $B \to \pi \pi $ see
ref.~\cite{cfms}. 

In conclusion, our analysis shows that measurements of CP asymmetries in
several channels may allow the extraction of the CP mixing phase and
to disentangle  SM and SUSY contributions to the CP decay phase.  
The golden-plated decays in this respect are $B \to \phi K_S$
and $B \to K_S \pi^0$ channels. The size of the SUSY effects is
clearly controlled by the the non-diagonal SUSY mass
insertions $\delta_{ij}$, which for illustration we have assumed to have the
maximal value compatible with the present experimental limits on 
$B^0_d$--$\bar B^0_d$ mixing.

A.M. acknowledges partial support from the EU contract ERBFMRX CT96 0090;
M.C., E.F. and  G.M. acknowledge partial support from EU contract 
 CHRX-CT93-0132.

\squeezetable
\begin{table}
 \begin{center}
 \begin{tabular}{cdddddd}  
 Process & $DE$ & $DE+CE$ &
 $DE+CE+$ & $DE+CE+$ &
 $DE+CE+$& $DE+CE+$\\
 &   &   &
 $CA$ & $CA+DE_{LR}+CE_{LR}$ &
 $DP+CP$ & $\overline{DP}+ \overline{CP}$\\
 \hline
&--&--&--&--&--&--\\
$B^0_d \to J/\psi K_S$&-0.03&0.1&0.1&0.1&0.1&0.1\\
&-0.008&0.02&0.02&0.04&0.02&0.02\\
\hline
&--&--&--&--&--&--\\
$B^0_d \to \phi K_S$&0.7&0.7&0.7&0.6&0.4&0.4\\
&0.2&0.2&0.2&0.1&0.1&0.09\\
\hline
&0.08&-0.06&-0.05&-0.02&-0.009&-0.01\\
$B^0_d \to K_S \pi^0$&0.7&0.7&0.6&0.6&0.4&0.4\\
&0.2&0.2&0.2&0.1&0.1&0.09\\
\hline
$B^0_d \to D^0_{CP} \pi^0$&0.02&0.02&0.02&0.02&0.02&0.02\\
\hline
&-0.6&0.9&-0.7&-2.&6.&4.\\
$B^0_d \to \pi^0 \pi^0$&0.3&-0.07&0.4&-0.4&-0.07&-0.06\\
&0.06&-0.02&0.09&-0.1&-0.02&-0.02\\
\hline
&-0.09&-0.1&-0.1&-0.3&-0.9&-0.8\\
$B^0_d \to \pi^+\pi^-$&0.02&0.02&0.03&0.09&0.8&0.4\\
&0.005&0.006&0.008&0.02&0.2&0.1\\
\hline
&0.03&0.04&0.05&0.1&0.3&0.2\\
$B^0_d \to D^+D^-$&-0.007&-0.008&-0.01&-0.02&-0.02&-0.02\\
&-0.002&-0.002&-0.002&-0.005&-0.006&-0.005\\
\hline
&0&0&0&0&0.&0.07\\
$B^0_d \to K^0 \bar{K}^0$&-0.2&-0.2&-0.2&-0.2&-0.09&-0.08\\
&-0.06&-0.05&-0.05&-0.04&-0.02&-0.02\\
\hline
&--&--&-0.2&-0.4&--&--\\
$B^0_d \to K^+K^-$&--&--&0.04&0.1&--&--\\
&--&--&0.01&0.03&--&--\\
\hline
&--&--&--&--&--&--\\
$B^0_d \to D^0\bar{D}^0$&--&--&-0.01&-0.03&--&--\\
&--&--&-0.003&-0.006&--&--\\
\hline
&-0.04&0.1&0.1&0.3&0.1&0.1\\
$B^0_d \to J/\psi \pi^0$&0.007&-0.02&-0.02&-0.03&-0.02&-0.02\\
&0.002&-0.005&-0.005&-0.008&-0.005&-0.005\\
\hline
&--&--&--&--&--&--\\
$B^0_d \to \phi\pi^0$&-0.06&-0.1&-0.1&-0.1&-0.1&-0.1\\
&-0.01&-0.03&-0.03&-0.03&-0.03&-0.03\\
 \end{tabular}
 \caption[]{\it{Ratios of amplitudes for exclusive $B$ decays. For each channel, 
whenever two terms  with  different CP phases contribute in the SM, 
we  give the ratio $r$ of the two amplitudes.
For each channel,
the second and third lines, where present, contain the ratios of SUSY to SM 
contributions for SUSY masses  of 250 and 500 GeV respectively.}}
 \protect\label{tab:ampli}
 \end{center}
 \end{table}
 
\begin{table}
 \begin{center}
 \begin{tabular}{ccccccc}  
 Incl. & Excl. & $\phi^{D}_{\rm SM}$ & $r_{\rm SM}$ & 
 $\phi^{D}_{\rm SUSY}$ & $r_{250}$ & $r_{500}$ \\
 \hline
 $b \to c \bar c s$ & $B \to J/\psi K_{S}$ & 0 & -- & $\phi_{23}$ & $0.03-0.1$
 &$0.008-0.04$ \\ \hline
 $b \to s \bar s s$ & $B \to \phi K_{S}$ & 0 & -- & $\phi_{23}$ & $0.4-0.7$ &
 $0.09-0.2$ \\ \hline
 $b \to u \bar u s$ &  & Penguin $0$ &  &  &  &
  \\
  &$ B \to \pi^{0} K_{S}$ &  & $0.009-0.08$ & $\phi_{23}$ & $0.4-0.7$ &
 $0.09-0.2$ \\ 
 $b \to d \bar d s$ &  & Tree $\gamma$ &  &  &  &
  \\ \hline
 $b \to c \bar u d$ &  & 0 &  &  &  &
  \\ 
  &$ B \to D^{0}_{CP} \pi^{0}$ &  & 0.02 & -- & -- &
 -- \\ 
 $b \to u \bar c d$ &  & $\gamma$ &  &  &  &
  \\ \hline
  & $B \to D^{+} D^{-}$ & Tree $0$ & $0.03-0.3$ &  & $0.007-0.02$ &
  $0.002-0.006$ \\ 
  $b \to c \bar c d$& &  &  & $\phi_{13}$ & &
  \\ 
  & $B \to J/\psi \pi^{0}$ & Penguin $\beta$ & $0.04-0.3$ & & $0.007-0.03$ &
  $0.002-0.008$ 
  \\ \hline
  & $B \to \phi \pi^{0}$ & Penguin $\beta$& -- & & $0.06-0.1$ &
  $0.01-0.03$ \\ 
  $b \to s \bar s d$& &  & & $\phi_{13}$ & &
  \\ 
  & $B \to K^{0} \bar{K}^{0}$ & {\it u}-Penguin $\gamma$ 
  & 0-0.07 & & $0.08-0.2$ &
  $0.02-0.06$ 
  \\ \hline
 $b \to u \bar u d$ & $B \to \pi^{+} \pi^{-}$ & Tree $\gamma$ 
 & $0.09-0.9$ & $\phi_{13}$ & $0.02-0.8$ &
 $0.005-0.2$ \\ 
 $b \to d \bar d d$ & $B \to \pi^{0} \pi^{0}$ & Penguin $\beta$ & $0.6-6$ 
 & $\phi_{13}$ & $0.06-0.4$ &
 $0.02-0.1$ \\ \hline
 & $B \to K^{+} K^{-}$ & Tree $\gamma$ & $0.2-0.4$ & & $0.04-0.1$ &$0.01-0.03$
  \\
 $b \bar d \to q \bar q$ & & & & $\phi_{13}$& & \\
 & $B \to D^{0} \bar D^{0}$ & Penguin $\beta$ 
 & only $\beta$ &  & $0.01-0.03$ &$0.003-0.006$ \\
 \end{tabular}
 \caption[]{\it{CP phases for B decays. $\phi^{D}_{SM}$ denotes the decay phase in
 the SM; for each
 channel, when two amplitudes with different weak phases are present, 
 one is given in the first row, the other in the last one
 and the ratio of the two  in the $r_{SM}$ column. $\phi^{D}_{SUSY}$
 denotes the phase of the SUSY amplitude, and the ratio of the SUSY to SM
 contributions is given in the $r_{250}$ and $r_{500}$ columns for the
 corresponding SUSY masses.}}
 \label{tab:results}
 \end{center}
 \end{table}
\end{document}